\documentstyle[aps,prl]{revtex}
\begin{document}
\draft
\twocolumn[\hsize\textwidth\columnwidth\hsize\csname@twocolumnfalse%
\endcsname

\author{Yang Ji, M. Heiblum, D. Sprinzak, D. Mahalu, Hadas Shtrikman}
\title{Phase Evolution in a Kondo correlated system}
\address{Braun Center for Submicron Research, Dept. of Condensed Matter
Physics\\ Weizmann Institute of Science, Rehovot, Israel 76100}
\maketitle

\begin{abstract}
The coherence and phase evolution of electrons in a mesoscopic system in the Kondo correlated regime were studied.
The Kondo effect, in turn, is one of the most fundamental many-body effects where a localized spin interacts with
conduction electrons in a conductor. Results were obtained by embedding a quantum dot (QD) in a \emph{double path}
electronic interferometer and measuring interference of electron waves. The Phase was found to evolve in a range
twice as large as the theoretically predicted one. Moreover, the phase proved to be highly sensitive to the onset
of Kondo correlation, thus serving as a new fingerprint of the Kondo effect.\\
\end{abstract}
]

\noindent The Kondo effect \cite{Kondo}, a many body phenomenon, first discovered in metals slightly doped with
magnetic impurities, has become one of the paradigms of strongly correlated systems. The effect, related to the
Anderson model \cite{Anderson}, results from an interaction of magnetic impurities with conduction electrons. In
1988 it was recognized \cite {Glazman}\cite{Ng} that the Kondo effect could also take place in a system of a spin
polarized quantum dot (QD) \cite{Meirav}, strongly coupled to an electron reservoir, thus forming a \emph{spin
singlet}. Indeed the effect was recently observed in a mesoscopic QD by Goldhaber-Gordon \textit{et al.}
\cite{DGG}, who were able to control \textit{in situ} many relevant parameters that affect Kondo correlation. Even
though the Kondo effect has been studied for some time, one of its most fundamental properties was never
experimentally verified: a phase shift $\pi /2$ experienced by scattering electrons from the spin singlet
\cite{KP1}\cite{KP2}. Recently, such a calculation was applied to a Kondo correlated QD with similar predictions
\cite{Delft}. We address this issue experimentally by combining a mesoscopic, Kondo correlated QD with a
\emph{double path} Aharonov - Bohm (AB) interferometer \cite {Yacoby}\cite{Shuster}. After verifying the
coherence, we proceeded to measuring the phase shift the traversing electrons through the dot experience. We find
an unexpected behavior of the phase shift, which evolved in a range  twice as large as that of the predicted one
\cite{Delft}. Moreover, we show that the phase evolution is extremely sensitive to the onset of Kondo correlation:
while the conductance of the QD varies slightly and in a monotonous fashion with such onset, the phase behavior is
radically different, serving as a distinct fingerprint of the Kondo effect.

The QD, serving here as a localized spin, is a small, confined, puddle of
electrons, with two tunnel barriers coupling it to two electron reservoirs.
The electrons in the puddle occupy a discrete ladder of energy levels, with
level energy $\varepsilon _{d}$, and an average energy separation $\Delta $
between non spin degenerate levels. A capacitively coupled metallic gate (%
\textit{plunger}) is used to tune the energy levels in the dot. Resonant tunneling between the two reservoirs
(through the tunnel barrier -- puddle -- tunnel barrier system) takes place and current flows when an energy level
is aligned with the Fermi level in the leads. When one of the electronic levels drops below the Fermi level in the
leads, it gets occupied and the number of electrons in the dot increases by one. However, due to the small
capacitance $C$ of the QD and the discreteness of the electronic charge $e$, an additional, classical,
\emph{charging energy} $U_{C}=e^{2}/2C$, is needed in order to add a single electron into the dot. Effectively,
the peak spacing is $U_{C}+\Delta$, where typically $U_{C}\gg\Delta$. At low temperature $T$ ($k_{B}T \ll U_{C}$,
with $k_{B}$ the Boltzmann constant), when the Fermi surface in the leads lays in the gap between levels, there is
a well-defined number of electrons in the QD and current does not flow. This is the well-known Coulomb Blockade
phenomenon, where periodic conductance peaks as a function of \textit{plunger} voltage are separated by almost
zero conductance valleys.

While Coulomb Blockade in QDs is a manifestation of the quantization of the electronic charge, the net spin of a
QD is also quantized (in units $h/2$, with $h$ the Plank constant). Generally, each energy state is spin
degenerate, namely, it can host two electrons: one with \emph{spin up} and one with \emph{spin down}. Hence, the
QD is \emph{spin polarized} (with nonzero net spin) when it contains an odd number of electrons, thus acting as a
'magnetic impurity'. Furthermore, when the QD is strongly coupled to the leads and the temperature is sufficiently
low, a spin singlet can form between the spin polarized dot and opposite spin electrons in the leads. This,
so-called, Kondo correlation leads to an enhancement of the conductance of the valley between the spin degenerate
peaks. The enhancement can be explained via an intuitive semi-classical approach. While the Coulomb Blockade is a
first-order quantum mechanical effect the Kondo effect is a second-order effect, involving two electrons
simultaneously. Having the top-most spin-degenerate energy level singly occupied with a spin up electron, this
electron can tunnel out to the right lead while another electron, with spin down, simultaneously tunnels in from
the left lead. This keeps the number of electrons in the dot constant but leads to a net spin flip in the dot.
This process is followed by the return to the dot of the spin up electron in the right lead accompanied by a
simultaneous jump of the spin down electron in to the right lead. The final result is a double-spin-flip of the
dot and net transfer of a spin down electron from the left lead to the right one - leading to net current transfer
\cite{Meir}\cite{Wingreen}. In an alternative view, for a dot well coupled to the leads, the spin up electron's
wavefunction leaks from the dot to the leads, keeping away the spin up electrons in the leads due to the Pauli
exclusion principle. The spin down electrons, on the other hand, can get very close to the polarized dot. This can
be modeled as an effective attractive potential for the spin down electrons near the dot, leading to the formation
of a lower energy spin singlet (Fig. 1A). Due to this attractive potential, which turns out to be most effective
for electrons at the Fermi energy, the impinging rate of spin down
electrons on the dot increases, leading effectively to an increased current. The \emph{%
effective transport density of states} develops a sharp peak at the Fermi energy (Fig. 1A), with a width
reflecting the binding energy of the singlet. This small binding energy is customarily described by a \emph{Kondo
temperature}, $T_{K}$ (some 1 K in our QDs). Hence, the Kondo enhanced valley conductance can be easily quenched by
increasing the temperature, applying a finite DC bias across the QD, or reducing the coupling strength between the
QD and the leads.

Ubiquitous conductance measurements in QDs have already demonstrated the
Kondo effect \cite{DGG}; however, they left the issue of coherence and phase
evolution in such system open. To address this issue, we adopted our
previously developed double-path electron interferometer with an embedded QD
in one of its paths \cite{Shuster} tuned to the Kondo regime (Fig. 1B). In
such system, smaller in overall dimensions than the \emph{phase breaking
length}, the current collected in the drain depends both on the magnitude, $%
t_{QD}$, and phase, $\phi _{QD}$, of the transmission coefficient of the QD.
Since the four base regions collect most of the back-scattered electrons,
the net transmission of electrons going from source (S) to drain (D), $t_{SD}
$, is a coherent sum of the transmissions via the two direct paths (dashed
lines in Fig. 1B). With $t_{ref}$, $\phi _{ref}$, belonging to the reference
arm, $t_{SD}=t_{ref}+t_{QD}$ (assuming $t_{left}=t_{QD}$). The collected
current in the drain, assuming a coherent system, is in turn $%
I_{SD}\propto \left| t_{SD}\right| ^{2}=\left| t_{left}\right| ^{2}+\left| t_{QD}\right| ^{2}+2\left|
t_{left}\right| \left| t_{QD}\right| cos(\phi _{ref}-\phi _{QD})$. Introducing a magnetic flux, $\Phi $, in the
area encompassed by the two paths, changes the relative phase of the reference arm via the AB effect (10),$\phi
_{ref}\longrightarrow \phi _{ref}+2\pi\Phi /\Phi _{0}$, with $\Phi _{0}=h/e$ the \emph{flux quantum}. This leads to
an oscillating periodic component in the current as a function of magnetic field $\propto $ $cos(\phi _{ref}-\phi
_{QD}+2\pi \Phi/\Phi _{0})$. The phase evolution in the QD can then be  directly deduced from the phase of the current
oscillation in the drain. Such experiments, which have already been carried out with QDs in the Coulomb Blockade
regime, lead to surprising results\cite {Shuster}. The measured phase of the QD was found to climb by $\pi $ as
each resonance swept through the Fermi level in the leads, as expected. In the valley between the peaks, however,
the phase experienced an unexpected abrupt $-\pi $ phase lapse --- a phenomenon still not understood.

The double path interferometer is similar to that used in Ref. 11 (Fig. 1B). It is fabricated in a two dimensional
electron gas (2DEG), embedded in a GaAs-AlGaAs heterostructure, with sub-micron metallic gates deposited on the
surface. The QD is made smaller than usual ($180nm\times 200nm$) to allow large energy level spacing, hence
allowing strong coupling of the dot to the leads, without overlapping of the energy levels. There are a few tens
of electrons in the dot with average level spacing $\Delta \cong 0.5meV$ and a charging energy $ U_{C}\cong
1.5meV$. Another \textit{barrier} gate was added in order to shut off the reference arm and allow measuring the
conductance of the bare QD. Measurements were done in a dilution refrigerator with base temperature $
T_{lattice}$$<50mK$ and an electron temperature $T_{electron} \sim 90mK$. Current collected in the drain was
measured with standard lock-in techniques, with an excitation voltage $10\mu V$ at $7Hz$, applied between the
source and the nearby base contacts. Four different samples, fabricated with different size QDs and in different
2DEG systems, were measured. All produced qualitatively similar results.

A \emph{Kondo-enhanced} valley, confined between a pair of conductance peaks, is usually identified by its higher
conductance and its quench either by an increased temperature, or by an applied finite DC bias across the QD, or
by reducing the coupling to the leads \cite{DGG}. Such an identification of the 'Kondo pair' was performed by
measuring the conductance of the QD after pinching off the reference arm with the \textit{barrier} gate. Four
conductance peaks as a function of plunger gate (P) voltage are seen in Figs. 1C and 1D. At the lowest temperature
and zero DC bias, the two center conductance peaks border a relatively high valley. These peaks are more closely
spaced and have peak heights near the quantum conductance, $e^{2}/h$. As the temperature was increased ($90\ldots
300$ mK, Fig. 1C) or the DC biased was raised ($0\ldots \pm 50 \mu V$, Fig. 1D), the central valley conductance
decreased, the peak heights decreased, and the peak spacing increased --- all characteristic of a Kondo correlated
pair \cite{DGG}. The conductance of the two outer valleys and peaks remained almost unchanged --- characteristic
of Coulomb Blockaded transport \cite{Meirav}.

Having identified the Kondo pair, we removed the \textit{barrier} gate voltage and formed the source and drain
point contacts, thus allowing two paths interference. The drain current, as a function of the \textit{plunger}
gate voltage, is shown in Fig. 2A. It resembles the conductance of the bare QD (seen in Fig. 1C), but the peaks
are shifted by some $80$ mV in \textit{plunger} gate voltage, their shape is distorted, and a large background
appears. The obvious reason for the shift is the change in the electrostatic influence of the \textit{barrier}
gate and the two point contacts forming the source and drain. Shape distortion can be accounted for by the
interference of the electron waves arriving from both of the interferometer arms. To perform interference, a
controlled phase difference between the two direct paths from S to D was then induced by the application of a
perpendicular magnetic field via the AB effect. The oscillation observed in the drain current (shown for example
in five different points in Fig. 2A and 2B), with period $3.5$ mT, immediately suggests coherency of transport in
the QD current, when the dot is tuned to peaks or a valley of conductance. The high visibility ($5-25\%$),
suggests that transport through the dot is highly coherent \cite{Yacoby}\cite {Shuster}. Assuming that the
transmission of the reference arm remains unaffected by the \textit{plunger} gate voltage, each trace of current
oscillation (like in Fig. 2B) provides two data points: an average visibility and phase, leading to the magnitude
and phase of  the transmission coefficient of the QD. The visibility and the phase are shown as a function of
\textit{plunger} gate voltage in Fig. 2C. The visibility mimics quite well the conductance of the QD shown in Fig.
1C, suggesting that the coherent transmission and the total transmission are qualitatively similar in the entire
energy range. While phase lapses, similar to those observed in Ref. 11, are observed between non-spin-degenerate
levels ( the outer valleys near $V_{P}=-280$ mV, $-210$ mV), the phase in the spin-degenerate levels
\textbf{evolves continuously and monotonously, with no sign of phase lapse.} The phase climbs through the spin up
peak and saturates at $\pi $ in the valley, thereafter it continues and climbs by anohter $0.5\pi $ through the
spin down peak. In other samples, not shown here, we find the climb through the spin down peak to be also close to
$\pi $. We believe that the total phase change across a Kondo pair varies from one sample to another because of
peaks overlapping
depending on their width and spacing. Hence, ideally, the phase seems to evolve by $%
\pi $ into the \textit{Kondo valley} and by $2\pi $ across the two spin degenerate levels. This is in
contradiction with the prediction of Gerland et al. \cite{Delft}, who predicted a $\pi /2$ and $\pi $ phase
change, respectively. It is worth noting that in our device, due to the relatively high temperatures of the
electrons, we could not reach the full Kondo enhancement, where the valley conductance reaches $e^{2}/h$
\cite{Delft}\cite {Wingreen}\cite{Schmid}. Still, phase behavior is markedly distinct and different than the
Coulomb Blockade regime. We believe it would stay invariant at lowest temperature.

Since the Coulomb Blockade regime can be easily approached by reducing the coupling strength between the QD and
the leads, such a procedure was followed with results shown in Fig. 3. As the coupling strength becomes gradually
weaker (from Fig. 3A to Fig. 3D), the phase evolves from a continuous increase across the two spin-degenerate
levels to a sudden appearance of a small phase lapse in the valley (even though the Kondo correlation  was still
apparent in that condition), increasing eventually to a full $-\pi $ phase lapse --- distinctive at Coulomb
Blockade transport. As suggested before, the narrowing of the peaks, leading to reduced overlapping between them,
indeed allowed a greater span of phase change, namely, a full accumulation of $\pi $  across each peak.

Similarly, increasing the temperature to an order of $T_{K}$ or applying a bias across the QD to an order of
$k_{B}T_{K}$ destroys Kondo correlation (Fig. 1). Note, however, that as the temperature increases, the visibility
tends to decrease because of the shorter dephasing length and phase smearing of the electrons in the
interferometer. While the visibility follows the behavior of the conductance shown in Fig. 1C, the phase
evolution, changes once again from that of a smooth and monotonic increase at low temperatures to that with a
small phase lapse in the valley as temperature increases. Phase lapse reaches  a full $-\pi $ drop at temperature
of about $1$ K (Fig. 4B). Similarly, the application of a small DC bias across the QD leads to a similar change in
the phase evolution, moving from a smooth increase to a phase lapse (Fig. 4D). In both cases Kondo correlation
ceases to exist at $T\sim T_{K}\sim 1$ K and $eV_{DC}\sim k_{B}T_{K}\sim100\mu eV$.

The results presented here demonstrate a clear fingerprint of Kondo correlation in a QD with spin degenerate
levels. As the correlation is being gradually suppressed, a gradual transition to the familiar behavior of the
phase in the Coulomb Blockade regime emerges. The total phase change as the two spin-degenerate levels cross the
Fermi energy in the leads is nearly $2\pi $ with a phase shift $\pi$ in the conduction valley. This does not agree
with the prediction by Gerland et al. \cite{Delft} of $\pi $ and $\pi /2$, respectively. Moreover, the phase
evolution approaches to be highly sensitive to the onset of Kondo correlation. While the conductance is only a
quantitative mark of the effect and has to be verified by changing temperature, voltage or coupling strength, the
phase evolution has a distinctive behavior, which abruptly changes as one weakens the Kondo correlation while
moving into the Coulomb Blockade regime.

\textbf{Figure Captions}

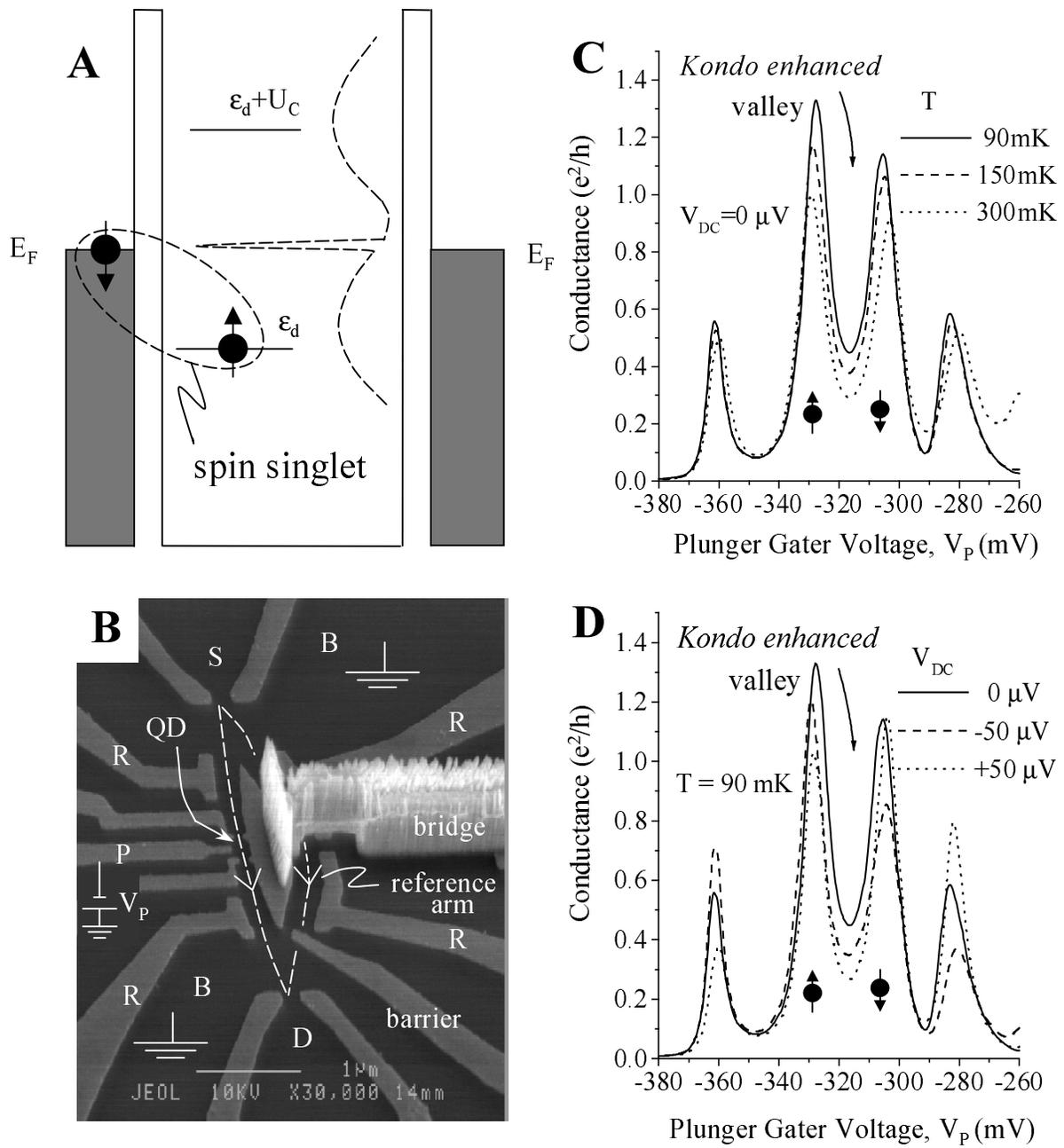
\textbf{Figure 1.}

Kondo correlated spin singlet, sample description, and characterization of a Kondo pair. (\textbf{A}) Energy
diagram of a quantum dot with Kondo correlation. The two spin-degenerate energy levels, $\varepsilon _{d}$ and
$\varepsilon _{d}+U_{C}$, are strongly coupled to the leads. At low temperature, a spin singlet is formed between
the localized electron in the QD and an opposite
spin electron in the leads, thus leading to a strong resonance in the \emph{%
transport density of states} (dashed line) centered at the Fermi level in the leads. (\textbf{B}) A top-view
scanning electron micrograph of the electronic double-path interferometer. The device is defined by negatively
biased metallic gates (lightly gray areas) deposited on the surface of a GaAs-AlGaAs heterostructure with high
mobility electrons $55$ nm below the surface (density $n=3\times 10^{11}cm^{-2}$, mobility $\mu =5\times
10^{5}cm^{2}V^{-1}s^{-1}$, measured at $1.5$ K). Three different regimes can be identified: source (S), drain (D),
and base (B). The source and the drain openings support only one transverse mode, thus producing a planar
electronic wavefront in the far field. The base regions collect back-scattered electrons to ensure that only the
two forward-propagating paths (broken white lines) reach the collector. A metallic air bridge, formed by special
lithographic techniques, is used to deplete the island between the two paths. A quantum dot (area $180nm\times
200nm$) is placed in the left arm, with both of its point contacts and the \textit{plunger} gate (P) individually
controlled. Reflectors R, confining the two paths, are used to deflect the electrons toward the drain. A
\textit{barrier} gate is used to cut off the reference (right) arm, in order to tune the bare dot to the Kondo
regime. (\textbf{C}) The effect of increasing temperature on the conductance with $V_{DC}=0\mu V$ across the QD.
As the temperature increases, the \emph(Kondo enhancemeant) of the central valley becomes smaller. (\textbf{D})
The effect of DC bias across the QD on the  conductance at $T=90$ mK. The central valley conductance decreases
with increasing DC bias.\bigskip

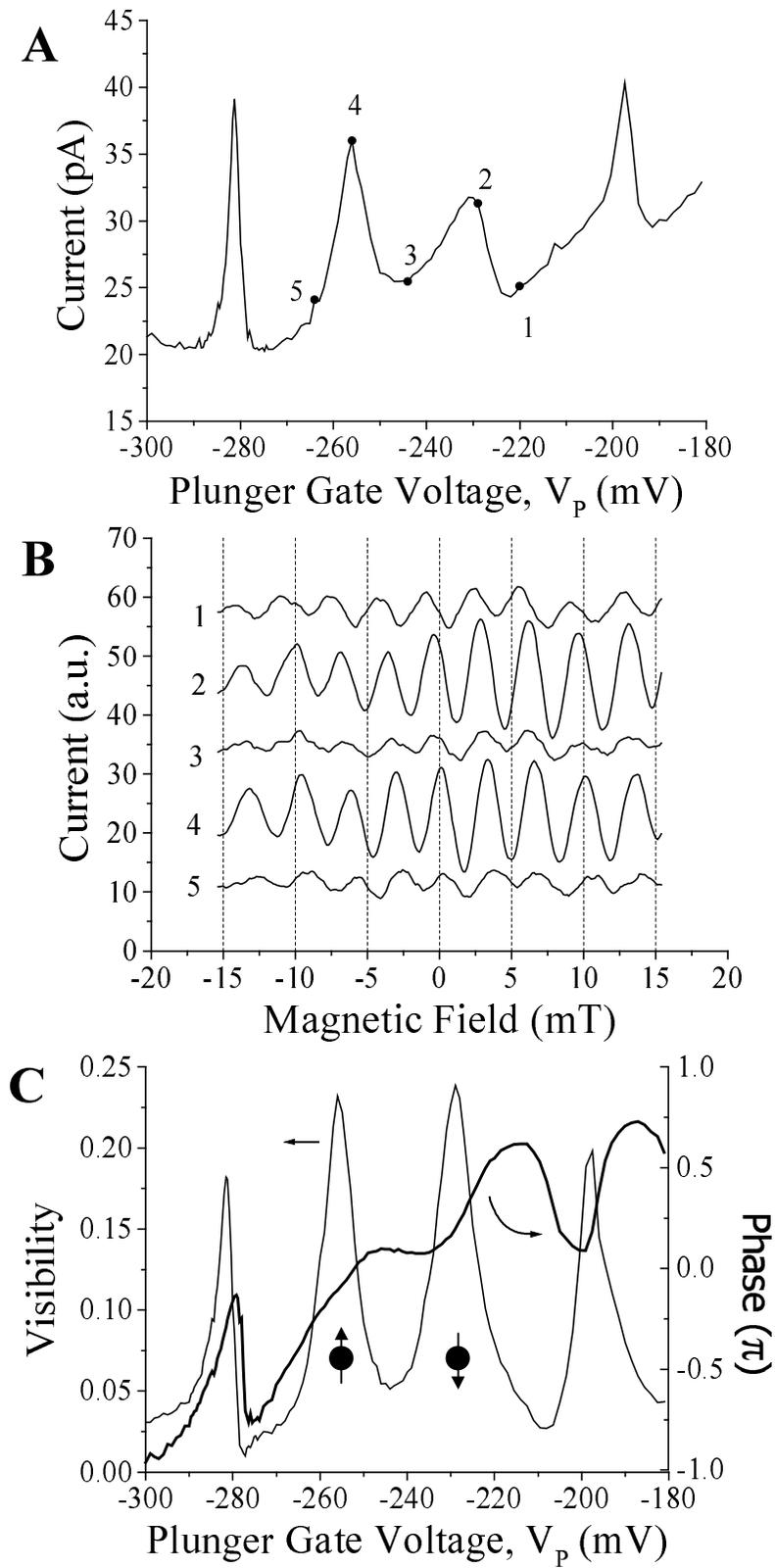
\textbf{Figure 2.}

Complex transmission coefficient of the quantum dot in the Kondo regime measured via the Aharonov-Bohm (AB)
effect. (\textbf{A}) Collector current as a function of \textit{plunger} gate voltage at zero magnetic field. This
current is very sensitive to magnetic field due to the AB effect,  as shown in (\textbf{B}), when a series of
interference patterns in the current collected by the drain (at specific points given in (A)), as a function of
magnetic field (shifted vertically), taken at specific positions noted in (A). Trace 3, measured at center of the
conductance valley between the Kondo pair, verifies coherency of electron transport in the Kondo valley.
(\textbf{C}) Qualitative behavior of the magnitude (shown by the visibility of AB oscillation) and absolute value
of the phase of the transmission coefficient of the QD --- both obtained from the AB oscillations. Note that the
phase evolves continuously through the Kondo pair while there is a phase lapse of $-\pi $ between the other
consecutive peaks.\bigskip

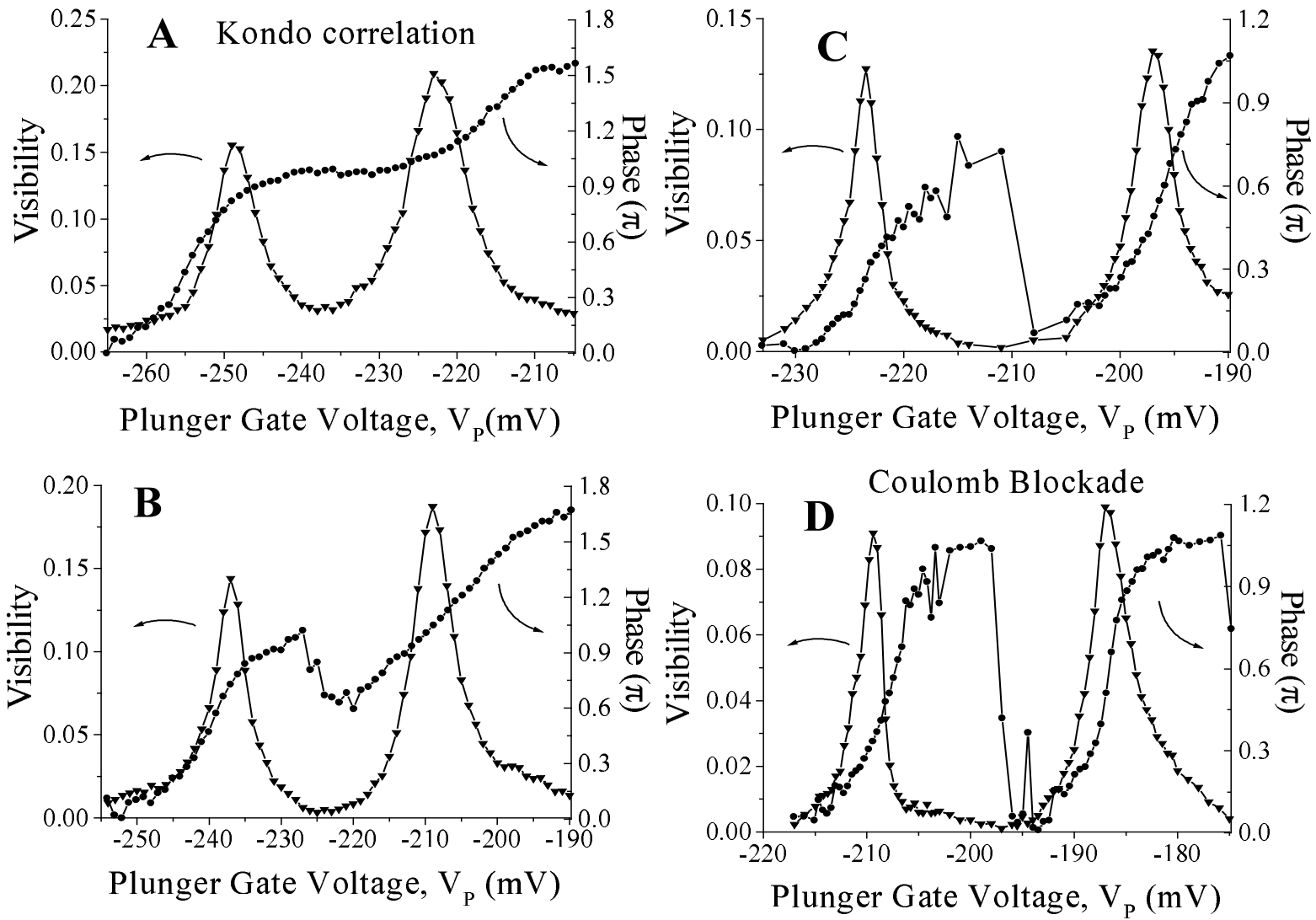
\textbf{Figure 3.}

Phase lapse is formed as the coupling between the quantum dot and the leads is gradually reduced. When the
coupling to the leads is strong (\textbf{A}), the phase evolution is continuous and the total phase change through
the spin-degenerate pair is about $1.5\pi$. As the coupling gets weaker (from (\textbf{A}) to (\textbf{D})), and
the QD enters the Coulomb Blockade regime, the familiar phase lapse \cite{Shuster} is recovered and a full phase
evolution of $\pi $ across each peak emerges. The peak position is shifting due to the influence of voltage
applied to the gates that formed the two point contacts  of the QD.\bigskip

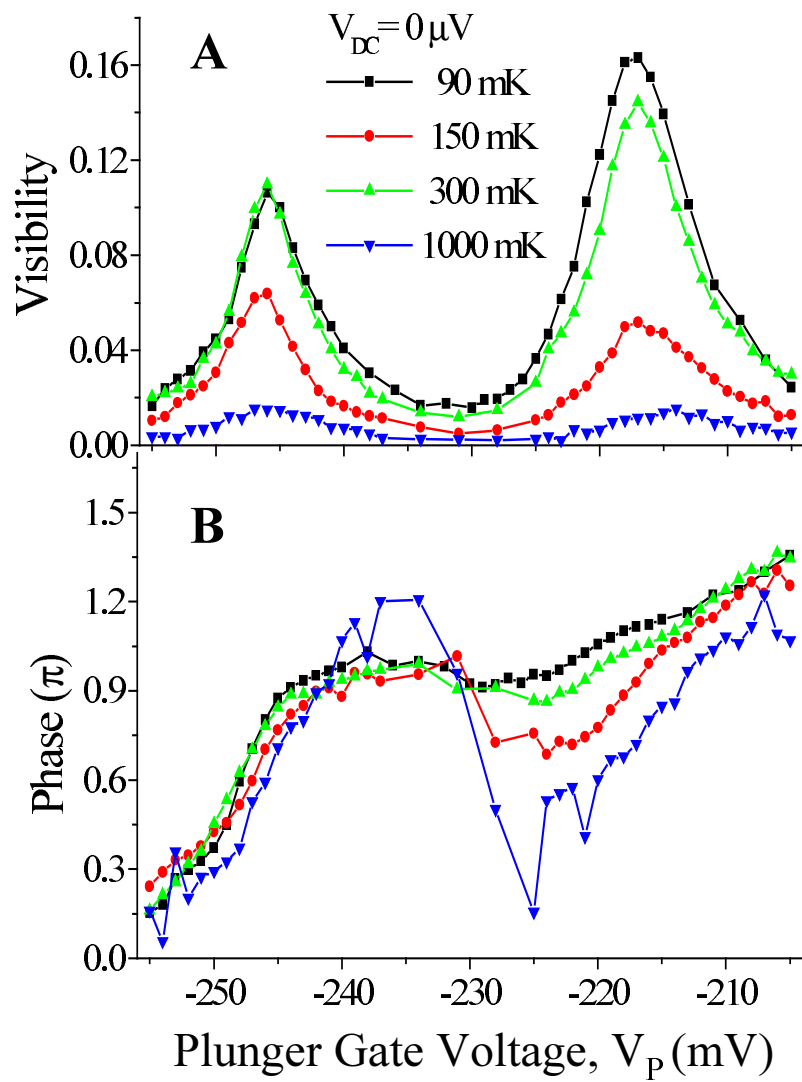
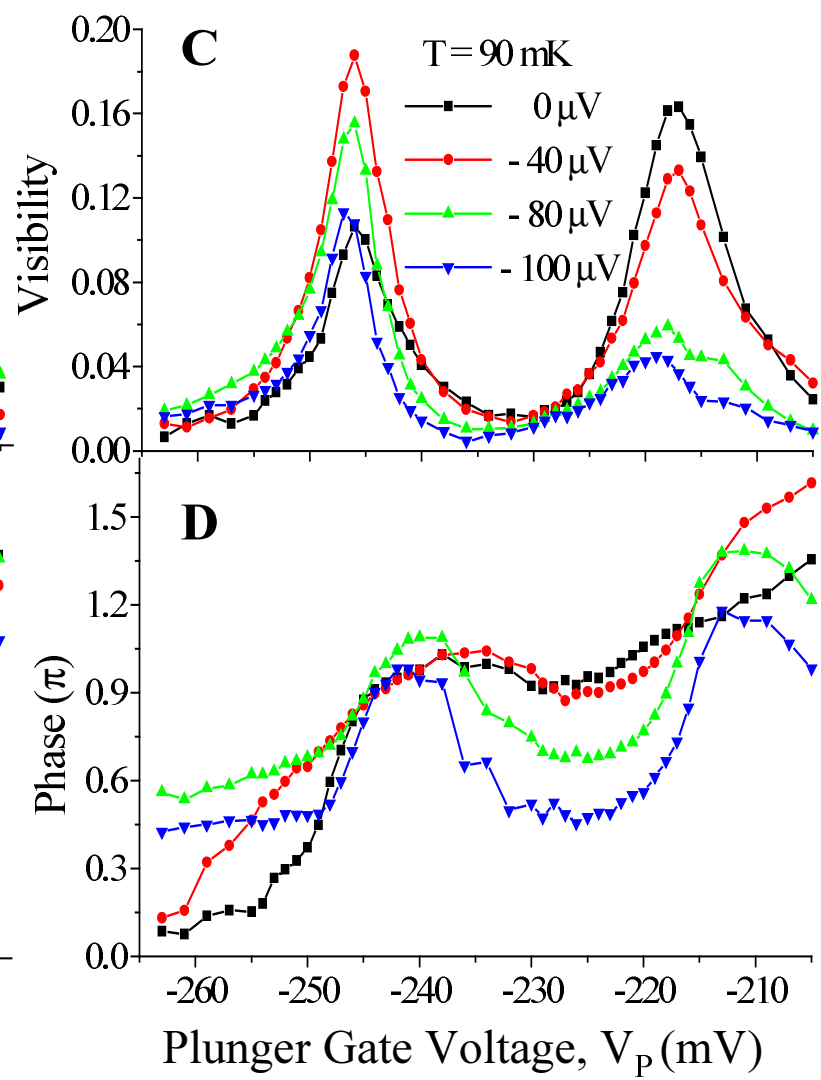
\textbf{Figure 4.}

The effect of temperature and an applied DC bias on the phase evolution. (\textbf{A}) Qualitative behavior of the
magnitude, and (\textbf{B}) phase evolution of the transmission coefficient as a function of \textit{plunger} gate
voltage, measured with increasing temperature. (\textbf{C}) qualitative behavior of the magnitude, and
(\textbf{D}) phase evolution of the transmission coefficient as a function of \textit{plunger} gate voltage,
measured at 90 mK with increasing DC bias between source and base. In both cases, the monotonous phase evolution
disappears at some point and a phase lapse is formed in the valley between the Kondo pair, as the QD approaches
the Coulomb Blockade regime.

\end{document}